# Magnetoelectric coupling in the cubic ferrimagnet $Cu_2OSeO_3$

Jan-Willem G. Bos[1],*, Claire V. Colin[2] and Thomas T.M. Palstra[2]

*School of Chemistry and Centre for Science at Extreme Conditions, University of Edinburgh, Edinburgh, EH9 3JJ, United Kingdom.*

*Solid State Chemistry Laboratory, Zernike Institute for Advanced Materials, University of Groningen, 9747 AG Groningen, The Netherlands.*

We have investigated the magnetoelectric coupling in the lone pair containing piezoelectric ferrimagnet $Cu_2OSeO_3$. Significant magnetocapacitance develops in the magnetically ordered state ($T_C$ = 60 K). We find critical behavior near $T_C$ and a divergence near the metamagnetic transition at 500 Oe. High-resolution X-ray and neutron powder diffraction measurements show that $Cu_2OSeO_3$ is metrically cubic down to 10 K but that the ferrimagnetic ordering reduces the symmetry to rhombohedral R3. The metric cubic lattice dimensions exclude a magnetoelectric coupling mechanism involving spontaneous lattice strain, and this is unique among magnetoelectric and multiferroic materials.

**Introduction**

Magnetoelectrics are materials in which an applied electric field can induce a magnetization or conversely where the application of a magnetic field leads to an induced electric polarization.[1,2] The magnetodielectric effect (changes in the dielectric constant at the magnetic ordering temperature or in a magnetic field) is often large close to a ferroic transition, which leads to large non-linear magnetoelectric (ME) couplings. Magnetoelectrics are usually materials that are magnetically ordered but not polar and show comparative modest linear ME coupling. Magnetoelectrics attracted significant interest in the 1970s and more recently with the enormous attention for multiferroic materials.[1-4] The earlier studies include work on $BaMnF_4$,[5-8] $Cr_2BeO_4$,[9] and $Gd_2(MoO_4)_3$,[10] while recently studied materials include SeCuO$_3$,[11] BiMnO$_3$,[12] EuTiO$_3$,[13] and CoCr$_2$O$_4$.[14, 15] Among these, ferromagnetic materials are of special interest as a large spontaneous magnetization M is considered to be favorable for large ME effects.[16] In spite of the significant theoretical and experimental interest there is no generic model to describe the observed dependence of the dielectric constant on spin structure and applied magnetic fields.[17] In modern multiferroics, two mechanisms describing the coupling between electric polarization and magnetic order have come to the fore: the spin-current model for spiral magnets and the exchange striction model for the RMn$_2$O$_5$ phases.[4] However, neither of these models makes quantitative predictions about the dielectric response. Other microscopic mechanisms include the coupling between long wavelength polar phonon modes and spin structure, as proposed for BaMnO$_4$,[7] and single ion effects.[18] Recent work on multiferroic materials has shown that anomalies in the dielectric constant that occur at the onset or with changes in the magnetic order are generally also associated with spontaneous lattice distortions.[19-23] This strongly suggests that the ME coupling proceeds via the lattice (atomic displacements).

Here, we present the results of our investigation into Cu$_2$OSeO$_3$, which shows significant ME coupling but does not have a spontaneous lattice distortion below T$_c$. Cu$_2$OSeO$_3$ is ferrimagnetic with a Curie temperature of 60 K and has a saturation magnetization of 0.50 µB/Cu.[24] At room temperature, it has the cubic space group P2$_1$3, which allows for piezoelectricity but not for a spontaneous polarization.[25, 26] Magnetic susceptibility measurements reveal a metamagnetic transition around 500 Oe between two ferrimagnetic states with different saturation magnetizations. Rietveld analysis of neutron powder diffraction data shows that the H = 0 magnetic structure is collinear ferrimagnetic with magnetic space group R3. This reduction in symmetry is required because ferrimagnetism is not symmetry allowed for cubic crystal structures.[27] However, high resolution synchrotron X-ray powder diffraction shows that there is no observable rhombohedral distortion below T$_c$. The crystal structure of Cu$_2$OSeO$_3$ therefore remains metrically cubic but the magnetic ordering lowers the crystal and magnetic symmetry to R3. In contrast to most studied magnetoelectric materials, the dielectric constant of Cu$_2$OSeO$_3$ is enhanced directly below the magnetic ordering transition, and shows positive magnetocapacitance (MC) near $T_C$. At lower temperatures the dielectric constant starts to decrease. Below $T_C$ a negative MC is observed, upon which the positive MC originating from the metamagnetic transition is superimposed.

To the best of our knowledge this is the first demonstration of ME coupling in a system where spontaneous lattice strain can be excluded, and as such is relevant to understand the microscopic coupling mechanisms in magnetoelectric and multiferroic materials. We use spontaneous strain to distinguish from the situation where strain is induced by an applied magnetic or electric field or by magnetic ordering. Induced strain participates in the ME coupling as both piezoelectric and piezomagnetic coupling are symmetry allowed, and may well be responsible for the observed magnetodielectric effects.

**Experimental**

Olive green polycrystalline samples of $Cu_2OSeO_3$ were prepared by standard solid state chemistry methods. CuO (99.999%) and $SeO_2$ (99.999%) were thoroughly mixed in a 2:1 ratio using mortar and pestle, and pressed into a pellet. The pellet was sealed in an evacuated quartz tube and heated to 600 °C over the duration of a day. After heating for 12 hours at 600 °C the sample was quenched, homogenized using mortar and pestle, pressed into a pellet and heated for 3 days at 600 °C with one more intermediate homogenization. Phase purity was confirmed by powder X-ray diffraction (PXRD). Variable temperature PXRD data (10 ≤ T ≤ 300 K) were collected using a Huber diffractometer with Mo $K_\alpha$ radiation. A 10 K dataset suitable for a full structure solution was collected on the ID31 diffractometer at the ESRF in Grenoble, France. Data were collected between 3 ≤ 2θ ≤ 50° and binned with a step size of 0.003°. No impurity phases were observed. The wavelength was 0.45620 Å. Neutron powder diffraction experiments were performed on the D20 instrument at the Institute Laue Langevin in Grenoble, France. Datasets were collected at 10 and 70 K using the low resolution high flux mode with a monochromator take-off angle of 44°. The neutron wavelength was 2.42 Å. Data were collected in the 10 ≤ 2θ ≤ 150° range in 0.1° increments. The GSAS suite of programs was used for Rietveld fitting of the powder diffraction data. Zero Field Cooled (ZFC) and Field Cooled (FC) magnetic susceptibilities in applied fields of 10 and 250 Oe were collected using a Quantum Design Magnetic Property Measurement System (MPMS). The field dependence of the magnetization was measured using a Quantum Design Physical Property Measurement System (PPMS) fitted with ACMS insert. Additional low field M(H) data (inset in Fig. 4b) at 5 K field were collected using the MPMS. The capacitance was measured in a commercial system (Quantum Design PPMS) using a home-made insert and an Andeen-Hagerling 2500A capacitance bridge operating at a fixed measurement frequency of 1kHz. Electrical contacts were painted using Ag-epoxy on a pressed pellet with capacitor geometry, typically 1*7*7 mm$^3$.

**Results**

*Structure*: The crystal structure of $Cu_2OSeO_3$ is depicted in Fig. 1a-b and is characterized by trigonal bi-pyramidal $CuO_5$, square pyramidal $CuO_5$ and tetrahedral $SeO_3$-lp (lp = lone pair) coordination polyhedra. The $CuO_5$ polyhedra share edges and corners while the $SeO_3$-lp polyhedra share corners with the $CuO_5$ polyhedra. The connectivity of the Cu ions is shown in Fig. 1b as are the local coordination environments. The $Cu^{2+}$ ions form a network of distorted tetrahedra whose corners are connected via linear Cu-Cu bridges. The solid lines indicate edge sharing $CuO_5$ and the open lines indicate corner sharing $CuO_5$ polyhedra. The Cu-Cu distances for edge-sharing $CuO_5$ are 0.18 to 0.25 Å shorter than the ones for corner sharing. The Cu coordination polyhedrons deviate significantly from ideal square pyramidal and trigonal bi-pyramidal, respectively (Fig. 1b). Bond valence sum calculations confirm the +2 oxidation state for the copper ions [BVS(Cu1)=2.06(2), BVS(Cu2)=2.02(2)].[28]

The temperature evolution of the lattice parameter (10 ≤ T ≤ 300 K) and the 10 K crystal structure were studied by PXRD (Fig. 2). All patterns were fitted using the space group $P2_13$.[25] No structural phase transitions were observed. The temperature dependence of the lattice constant is shown in Fig. 3. The data have been scaled using the ID31 dataset at 10 K. The 300 K cell constant ($a$ = 8.9235(2) Å) is in good agreement with the literature value ($a$ = 8.925(1) Å).[25] The solid line is a fit to a(T) = $a_0$ + Acoth($\theta$/T), which is an approximation to the bare thermal expansion, due to thermal vibrations, of a solid as derived in Ref. [29]. ($\theta$ equals half the Einstein temperature). Deviations from this temperature dependence signal the occurrence of anomalous lattice strain. No deviations were observed and $Cu_2OSeO_3$ has a conventional thermal expansion due to lattice vibrations. A comparison of bond lengths and angles at 300 and 10 K does not reveal any significant changes (the 10 K crystallographic coordinates are given in Table 1), and further confirms the absence of any magneto-structural coupling or polar structural distortions.

*Magnetism*: The temperature dependences of the ZFC and FC magnetic susceptibilities, and the inverse ZFC susceptibility, collected in H = 250 Oe, are shown in Fig. 4a. The susceptibility diverges just below 60 K. Above 100 K, the susceptibility follows the Curie-Weiss law, and a fit (solid line) gives a Curie constant of 0.23(1) emu mol $Cu^{-1}$ $Oe^{-1}$ $K^{-1}$ and Weiss temperature of +69(2) K. The positive Weiss temperature indicates the presence of dominant ferromagnetic interactions, in agreement with the observed divergence of the susceptibility. The experimental effective moment (1.36 $\mu_B$/Cu) is lower than the expected spin-only value for S = ½ of $Cu^{2+}$ (1.73 $\mu_B$). This reduction is not unusual for $Cu^{2+}$ in metal oxides, e.g. in CuO and $La_2CuO_4$ the moment is reduced to ~50-70% of the spin only value.[30, 31] The field dependence of the magnetization is shown in Fig. 4b. The magnetization saturates in small applied fields and has a saturation value of 0.50 $\mu_B$/Cu at 5 K. This is exactly half the expected saturation moment for a S = ½ ferromagnet, and suggests a simple collinear ferrimagnetic alignment with 3 majority and 1 minority spins. A change of slope can be noticed around 500 Oe (Fig. 4b). This signals the presence of a metamagnetic transition with a small amount of magnetic hysteresis (insets to Fig. 4b). Extrapolating the low field magnetization suggests a saturation moment around 0.25(5) $\mu_B$/Cu at 2 kOe. The inset to Fig. 4a shows the low temperature ZFC and FC curves in 10 Oe, confirming the ferrimagnetic state even in small applied fields.

*Neutron powder diffraction*: Long range magnetic order was confirmed by the observation of magnetic intensities on the (110) and (201) reflections in the 10 K diffraction pattern (inset to Fig 5a). The magnetic cell is identical to the crystallographic one, and magnetic symmetry was used to construct possible magnetic models. The only possible magnetic space group (MSG) based on the crystal structure is $P2_13$. This MSG, however, does not allow for ferromagnetic or ferrimagnetic magnetic ordering. In fact, no cubic MSG allows for ferromagnetic ordering, and a symmetry lowering is therefore required.[27] Possible crystallographic subgroups are R3 and $P2_12_12_1$, and ferrimagnetic structures are possible in MSGs R3 (m // 3) and $P2_12_1'2_1'$ (m // $2_1$). Rietveld refinement revealed that models with anti-parallel sublattices based on the Cu1 and Cu2 site from the $P2_13$ structure gave the best fits. This corresponds to a magnetic structure with 12 majority and 4 minority spins. The R3 solution is shown in Fig. 5b. The Cu moments refine to $m_x=m_y=m_z=0.35(3)$ $\mu_B$ for Cu1 and $m_x=m_y=m_z= -0.35(2)$ for Cu2, yielding a moment of 0.61(5) $\mu_B$ per copper ($wR_p$=1.5%, $R_F^2$= 8.86). The reduced ordered moment (the expected spin-only value is 1 $\mu_B$) is common in copper oxides, e.g. CuO has m=0.68 $\mu_B$/Cu.[30] For the $P2_12_1'2_1'$ model m=$m_x$=0.61(5) $\mu_B$, and an identical goodness of fit was obtained. This ferrimagnetic arrangement corresponds to a saturation moment of 0.3 $\mu_B$/Cu in good agreement with the extrapolated low field magnetization (Fig. 4). A comparison of the magnetic structure (Fig. 5b) and the crystal structure (Fig. 1b) is of some interest. The Kanamori-Goodenough rules predict ferromagnetic exchange interactions for edge-sharing $CuO_5$ polyhedra (solid lines) and antiferromagnetic exchange for corner-sharing (open lines). The experimental magnetic structure is largely consistent with this. All exchange interactions are satisfied within the Cu4 tetrahedra but the coupling between tetrahedra is not as expected based on the Kanamori-Goodenough rules.

*Dielectric constant*: The temperature dependence of the dielectric constant is shown in Fig. 6a. Immediately below 60 K, the dielectric constant is enhanced at the emergence of long range order. The enhancement is noteworthy since a reduction in dielectric constant is more common, irrespective of the type of magnetic order. For example, BiMnO$_3$ (FM) SeCuO$_3$ (FM) and YMnO$_3$ (AFM) all show a reduction below the magnetic ordering transition.[11, 12, 32] Upon further cooling the dielectric constant decreases, and below 20 K is lower than the extrapolated lattice contribution (see below). This could reflect the complex temperature evolution of the magnetic order parameter in zero applied field. The lattice contribution to the dielectric constant was fitted (100 ≤ T ≤ 200 K) using the expression for the lattice thermal expansion. The Einstein temperature was fixed at 316 K, and the fit results are given in Fig. 6a. Subtraction of the lattice contribution allows for the analysis of the critical behavior in the vicinity of the magnetic ordering transition. The dielectric constant (minus the lattice contribution) can be fitted with:[32]

$$|\varepsilon_s(T) - \varepsilon_s(T_c)| = A|(T/T_c) - 1|^{1-\alpha} \qquad (1)$$

For $T_c$ = 59.5 K, the fit shown in Fig. 6b leads to equal slopes before and after the transition. The critical exponent α can be estimated from the slope (1-α) and α ≈ 0.3. The critical exponent for AFM FE YMnO$_3$ is 0.25.[32]

The magnetocapacitance: $MC = (C(H) - C(0))/C(0)$ was measured for temperatures between 5 and 60 K and for fields between 0 and 8 Tesla (Fig 7a-b). At low temperatures and fields, the magnetocapacitance is dominated by a peak at the metamagnetic transition. The peak position is hysteretic revealing that the metamagnetic transition is of first order (inset to Fig. 7b). At higher fields the magnetocapacitance decreases gradually with field. In contrast, in the critical region, the magnetocapacitance is smooth, positive and has a convex curvature. This curvature is also observed for ferromagnetic SeCuO$_3$ and BiMnO$_3$ but in those cases the magnetocapacitance is negative.[11, 12] Antiferromagnetic materials, such as YMnO$_3$ and TeCuO$_3$, in contrast, have concave negative magnetocapacitances.[11, 33] The magnetocapacitance can be described in a phenomenological manner using:

$$MC = MC_0 + A|H - H_c|^{\alpha} \quad (2)$$

at low temperatures, where $\alpha \approx -0.49$ and $H_c$ is the field of the metamagnetic transition, and

$$MC = MC_0 + BH^{\beta} \quad (3)$$

near the magnetic transition. In the intermediate regime these two functions can be added to fit the magnetocapacitance, and A and B are a measure of the weights of the long range ordered magnetic and critical contribution, respectively. The temperature dependences of $MC_0$, A, B and $\beta$ are given in figure 7c. The critical contribution can be seen to peak in the vicinity of the magnetic transition, while the long range magnetic contribution is most significant at lower temperatures, and vanishes at $T_c$. The exponent $\beta$ varies between ~0.3 near $T_C$ and approaches ~1 at low temperatures. The field independent term ($MC_0$) gradually decreases with temperature, and has a local maximum just above the magnetic transition.

**Discussion**

The coupling between magnetic and polar order parameters in multiferroic materials attracts much interest but little is known about the microscopic origin. From Landau theory, the dominant symmetry unrestricted coupling terms are non-linear terms in the free energy such as $M^2P^2$ or $L^2P^2$, with L the antiferromagnetic sublattice magnetization, M the magnetization and P the electric polarization. Multiferroics are expected to show strong coupling as the non-linear terms are large in the vicinity of a phase transition. However, no experimental realization of a material with large P and M is known. Materials like $BiMnO_3$, $SeCuO_3$ and $Cu_2OSeO_3$ with large M but no spontaneous electric polarization show nevertheless significant magnetoelectric coupling. This suggests that the large ordered magnetic moment is important for the coupling. Here, the coupling may operate via an induced polarization by the applied electric field used in the measurements. Alternatively, it has also been proposed that the magnetoelectric coupling proceeds via lattice strain.

$Cu_2OSeO_3$ is an interesting model system. This is mainly due to the fact that our structural studies show no measurable structural distortion occurs down 10 K. This excludes the possibility that the magnetoelectric coupling proceeds via a spontaneous lattice distortion below the magnetic ordering temperature. The dielectric constant initially increases below the magnetic ordering temperature. This is followed by a decrease and below about 20 K the dielectric constant is smaller than the extrapolated lattice contribution. This unusual temperature dependence is probably related to the different temperature dependence of the ferro- and antiferromagnetic order parameters, $M(T)$ and $L(T)$ respectively. Critical behavior similar to that observed for $YMnO_3$ is observed near the transition.[32] The initial increase in dielectric constant and positive magnetocapacitance near the magnetic transition are unusual. Most studied ferromagnetic and antiferromagnetic materials have negative magnetocapacitance and show a decrease in dielectric constant. Magnetization and magnetocapacitance measurements reveal a metamagnetic transition around 500 Oe. The metamagnetic transition shows up as a large positive peak in the magnetocapacitance measurements, which becomes stronger at lower temperatures, and is superposed on a decreasing capacitance. The high field magnetic state is consistent with a simple collinear ferrimagnetic arrangement with 3 majority and 1 minority $S=1/2$ spin, leading to a saturation moment of 0.5 $\mu_B$/Cu. The magnetic structure in zero magnetic field was determined from neutron powder diffraction and is also collinear ferrimagnetic but with a reduced copper moment (0.61(5) $\mu_B$). Ferrimagnetic structures are incompatible with cubic symmetry, and the crystal and magnetic structure below $T_c$ are therefore described in R3.

The absence of lattice strain indicates that linear magnetoelectric coupling effects may be important as expected for non-multiferroic materials. The magnetic space group R3 allows for piezoelectric coupling, and for both a linear magnetoelectric effect and coupling via a piezomagnetic effect. Thus, $Cu_2OSeO_3$ is a unique example of a metrically cubic material that allows linear magnetoelectric coupling as well as piezoelectric and piezomagnetic coupling. Further measurements are needed to find out which linear coupling terms of the magnetoelectric tensor dominate.

**Acknowlegdements**

JWGB acknowledges the Royal Society of Edinburgh for financial support, and EPSRC for provision of the beam time at the ESRF and ILL. Dr. Andy Fitch, Dr. Paul Henry and Dr. Simon Kimber are acknowledged for help with data collection. CC acknowledges the EU STREP project COMEPHS under contract No. NMPT4-CT-2005-517039. TP acknowledges stimulating discussions with Umut Adem, Beatriz Noheda, and Maxim Mostovoy.

**Figure Captions**

Fig. 1 (color online) (a) Polyhedral representation of the crystal structure of $Cu_2OSeO_3$ with $SeO_3$lp polyhedrons omitted. (b) Connectivity of the Cu ions and local coordination environments of Cu1 and Cu2. The labeling of atoms is consistent with that used in Table 1.

Fig. 2 (color online) Observed (crosses), calculated (solid line) and difference PXRD profiles for $Cu_2OSeO_3$ at 160 K (laboratory data) and at 10 K (synchrotron data). The markers correspond to the Bragg positions of $Cu_2OSeO_3$.

Fig. 3 Temperature evolution of the cubic lattice constant.

Fig. 4 (a) The temperature dependences of the ZFC and FC magnetic susceptibilities in an applied field of 250 Oe. The Curie-Weiss fit to the inverse ZFC susceptibility is shown. The inset shows the ZFC and FC susceptibilities measured in 10 Oe. (b) The field dependence of the magnetization at 5, 25, 50 and 75 K. The insets illustrate the high field behavior, the metamagnetic transition and the associated small magnetic hysteresis.

Fig. 5 (color online) (a) Observed (crosses), calculated (full line) and difference neutron powder diffraction Rietveld profiles for $Cu_2OSeO_3$ at 10 K. The inset shows the 10 K dataset fitted with the structural model obtained at 70 K. The magnetic reflections are indexed. The markers correspond to the Bragg positions. (b) Representation of the R3 magnetic structure. White circles correspond to Cu1 sites while blue ones correspond to Cu2.

Fig. 6 (a) The temperature dependence of the dielectric constant. The solid line is a fit. (b) critical behavior of the magnetic contribution to the dielectric constant.

Fig. 7 (color online) (a) Magnetocapacitance in low applied fields illustrating the anomaly that occurs at the metamagnetic transition. The inset shows the magnetic H-T phase diagram. (b) Fits to the field dependence of the magnetocapacitance (see text) at temperatures below (T = 25 K) and at the magnetic transition (T = 60 K). (c) Temperature evolution of the fitting constants $MC_0$, A, B and β (see text).

Table 1. Atomic parameters for $Cu_2OSeO_3$ at 10 K obtained from Rietveld fitting of the ID31 data. Goodness of fit statistics: $\chi^2$ = 6.6, $wR_p$ = 10.6%, $R_p$ = 6.8%, $R_F^2$ = 2.4%. Space group $P2_13$, $a$ = 8.91113(1) Å.

|     |     | $x$ | $y$ | $z$ | $U_{iso}$ (Å$^2$) |
| --- | --- | --- | --- | --- | --- |
| Cu1 | 4$a$  | 0.88557(6)  | =$x$        | =$x$         | 0.00105(5)  |
| Cu2 | 12$b$ | 0.13479(6)  | 0.12096(6)  | -0.12733(6)  | 0.00105(5)  |
| O1  | 4$a$  | 0.0103(3)   | =$x$        | =$x$         | 0.0022(4)   |
| O2  | 4$a$  | 0.7619(4)   | =$x$        | =$x$         | 0.0022(4)   |
| Se1 | 4$a$  | 0.45993(5)  | =$x$        | =$x$         | 0.00075(5)  |
| Se2 | 4$a$  | 0.21223(5)  | =$x$        | =$x$         | 0.00075(5)  |
| O3  | 12$b$ | 0.2306(3)   | 0.5159(3)   | -0.0301(3)   | 0.0034(4)   |
| O4  | 12$b$ | 0.2731(3)   | 0.1872(3)   | 0.0331(3)    | 0.0034(4)   |

Fig. 1a

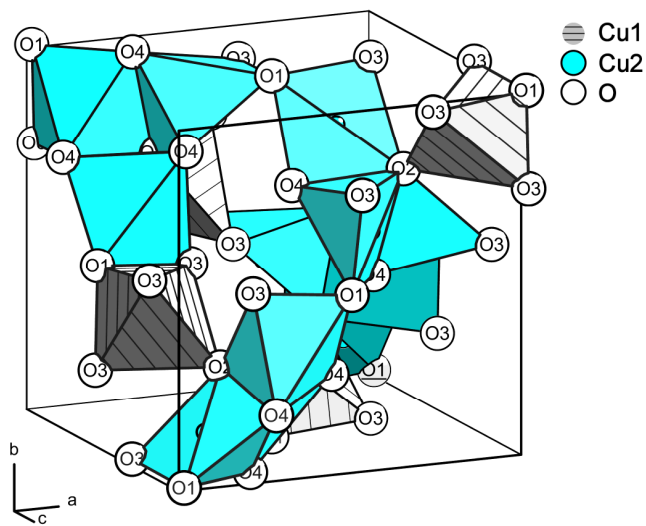

Fig. 1b

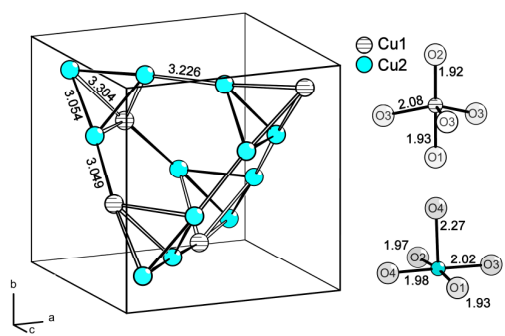

Fig. 2

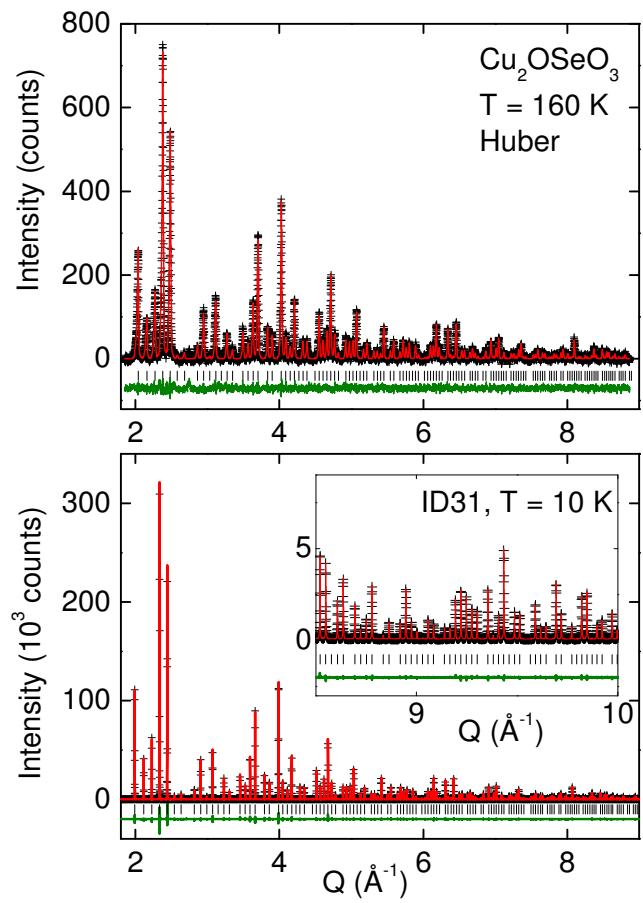

Fig. 3.

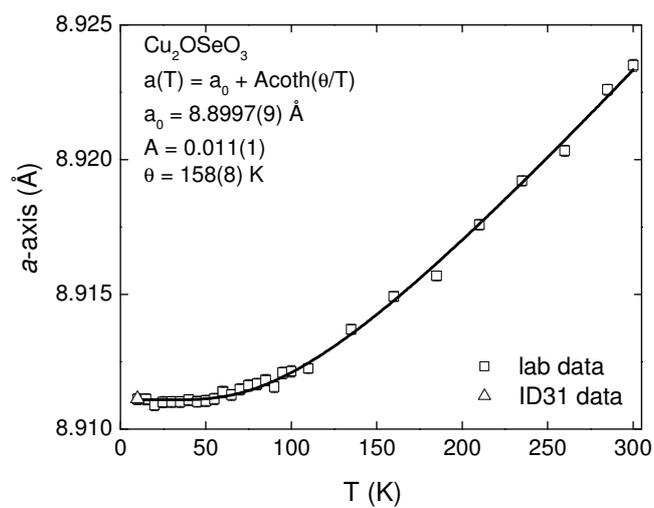

Fig. 4a

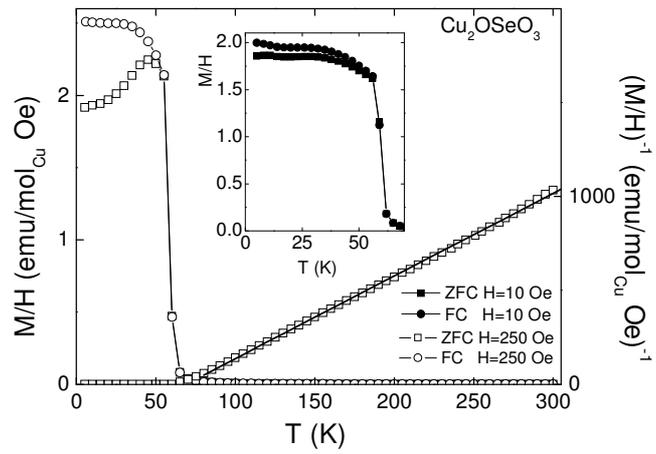

Fig. 4b

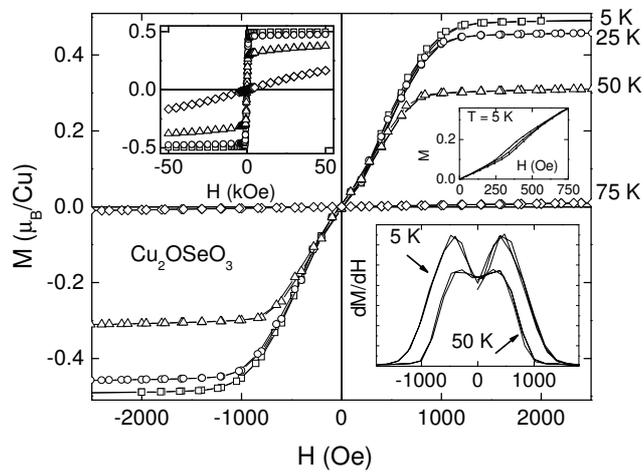

Fig 5a

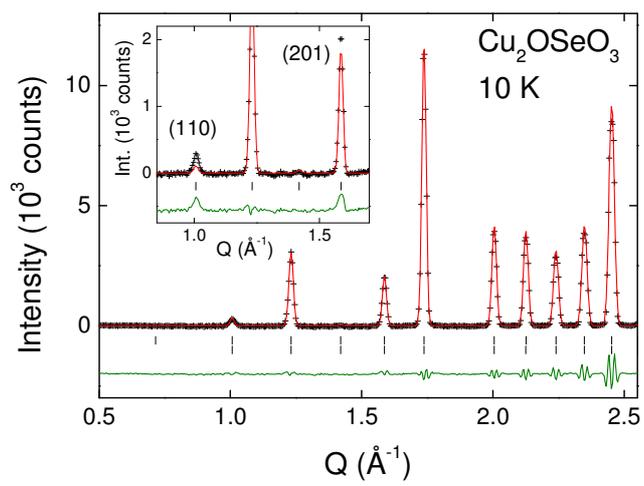

Fig. 5b

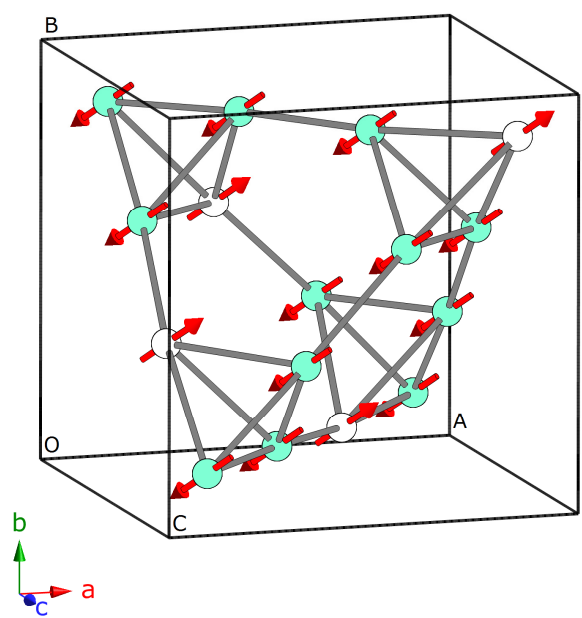

Fig. 6a

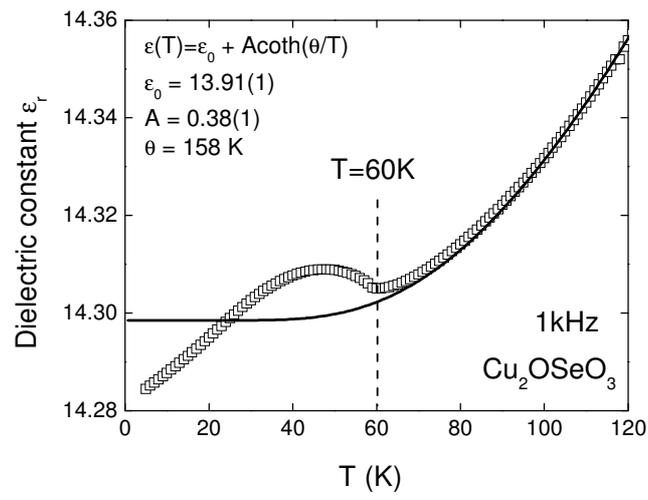

Fig. 6b

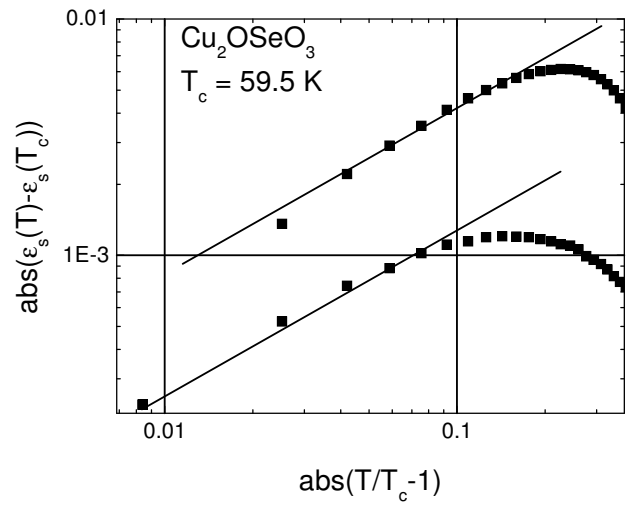

Fig. 7a

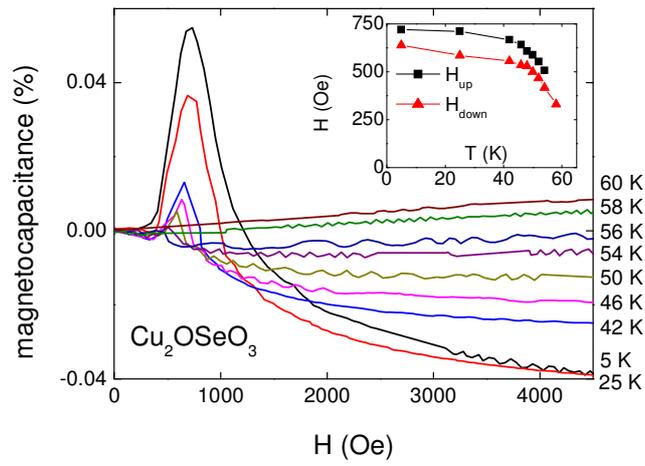

Fig. 7b

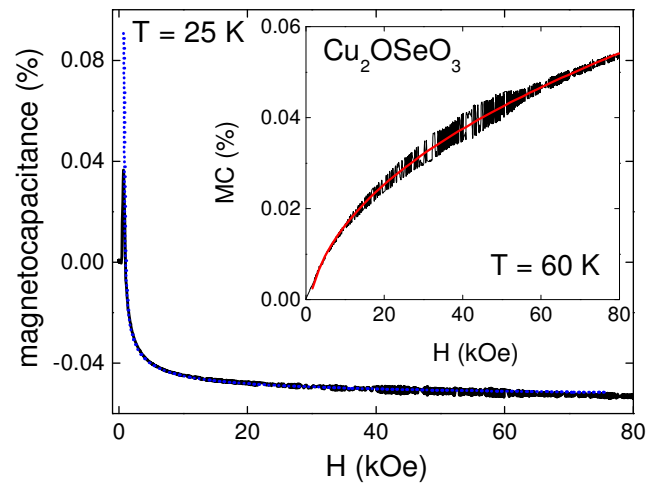

Fig. 7c

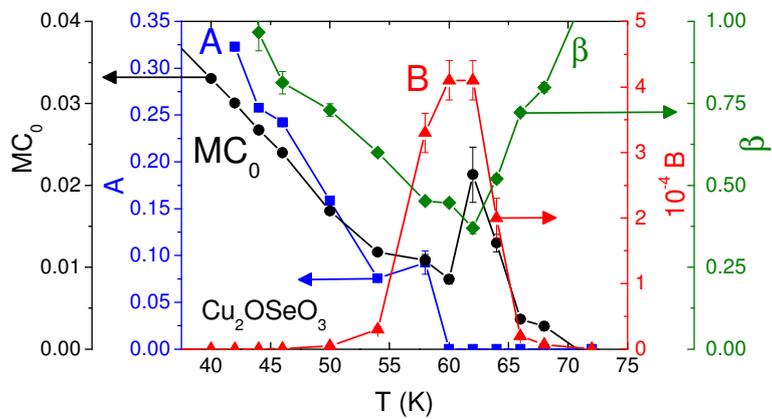